\documentclass[aps,prx,twocolumn,showpacs,floatfix,notitlepage]{revtex4-1}
\usepackage[latin9]{inputenc}
\usepackage{color}
\usepackage{verbatim}
\usepackage{units}
\usepackage{amsmath}
\usepackage{amssymb}
\usepackage{graphicx}
\usepackage{graphicx}  
\usepackage{epstopdf}
\usepackage{bm}        
\usepackage{amsmath}
\usepackage{amssymb}   
\usepackage{array}
\usepackage[caption=false]{subfig}
\usepackage[bookmarks=false,linkcolor=blue,urlcolor=blue,colorlinks,citecolor=blue]{hyperref}
\usepackage{dsfont}
\usepackage{appendix}

\makeatletter
\@ifundefined{textcolor}{}
{%
 \definecolor{BLACK}{gray}{0}
 \definecolor{WHITE}{gray}{1}
 \definecolor{RED}{rgb}{1,0,0}
 \definecolor{GREEN}{rgb}{0,1,0}
 \definecolor{BLUE}{rgb}{0,0,1}
 \definecolor{CYAN}{cmyk}{1,0,0,0}
 \definecolor{MAGENTA}{cmyk}{0,1,0,0}
 \definecolor{YELLOW}{cmyk}{0,0,1,0}
}

\usepackage{dcolumn}
\usepackage{bm}
\usepackage{setspace}
\usepackage{fancybox}
\usepackage{listings}
\usepackage{url}
\usepackage{pdfsync}
\usepackage{float}

\usepackage{subfig}
\begin{document}
\title{Generalized parafermions and non-local Josephson effect in multi-layer systems}

\author{Hiromi Ebisu,$^{1}$ Eran Sagi,$^{2}$ Yukio Tanaka,$^{1}$ and Yuval Oreg$^{2}$
}
\affiliation{$^1$~Department of Applied Physics, Nagoya University, Nagoya 464-8603, Japan \\
$^2$~Department of Condensed Matter Physics, Weizmann Institute of Science, Rehovot, Israel 76100\\
}

\date{\today}
\begin{abstract}
We theoretically investigate the effects of backscattering and superconducting proximity terms between the edges of two multi-layer fractional quantum Hall (FQH) systems. While the different layers are strongly interacting, we assume that tunneling between them is absent.
Studying the boundaries between regions gapped by the two
mechanisms in an $N$-layer system, we find $N$ localized zero-mode operators
realizing a generalized parafermionic algebra.  We further propose an
experiment capable of probing imprints of the generalized parafermionic bound states.
This is done by coupling different superconducting contacts to different
layers, and examining the periodicity of the Josephson effect as a function
of the various relative superconducting phases. Remarkably, even if we apply
a phase difference between the superconductors in one layer, we induce a
Josephson current at the other layers due to inter-layer interactions.
Furthermore, while the Josephson effect is commonly used to probe only
charged degrees of freedom, the possibility of independently controlling the
superconducting phase differences between the layers allows us to find
imprints of the neutral modes of the underlying multi-layer system. In
particular, we propose two configurations, one of which is capable of
isolating the signal associated with the charge modes, while the other probes
the neutral modes.
\end{abstract}
\pacs{73.43.-f,74.50.+r,05.30.Pr,71.10.Pm}


\maketitle
\thispagestyle{empty}
\section{Introduction}
The fractional quantum Hall (FQH) effect \cite{Tsui1982} was historically the first known realization of topologically ordered phases of matter \cite{Wen1990a,Wen1990b}. Similar to the much simpler integer quantum Hall (IQH) states, these remarkable states generally exhibit gapless edge modes, topologically protected by the gapped bulk. However, as opposed to IQH states, the bulk excitations generally carry fractional charges and anyonic statistics.
\par
Of particular interest are non-Abelian FQH states, whose bulk excitations are anyons characterized by non-Abelian braiding statistics. The non-local nature of operations generated by braiding the non-Abelian anyons makes these states promising candidates as platforms for quantum information processing. Indeed, it has long been suspected that the plateaus observed at filling factors $\nu=5/2$ and $\nu = 12/5$ are described by the non-Abelian Moore-Read \cite{Moore1991} and $k=3$ Read-Rezayi \cite{Read1999} states, respectively. However, a conclusive experimental confirmation has yet to be reported \cite{Dolev2008,Willett2009,Lin2012,Willet2010,Bid2010,Stern2010,Venkatachalam2011,Rhone2011,Tiemann2012,Chickering2013,Radu2008,Baer2014,Stern2012,Lin2012}.
\begin{figure}[h]
 \begin{center}
 \includegraphics[width=.45 \textwidth]{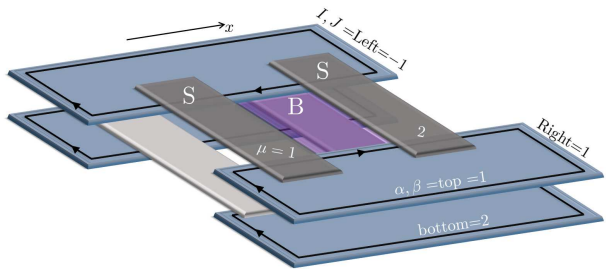}
 \end{center}

\caption{A schematic depiction of the two double-layer fractional quantum-Hall systems we study. Two superconducting contacts, denoted by the letter `S', are placed on each layer (these are indicated by the dark gray regions on the top layer and the light gray regions on the bottom layer). Between the two superconductors, the right and left moving edge modes are coupled by backscattering terms (these are indicated by the purple region and the letter `B'). The notations introduced in the figure are used throughout the paper: $\alpha$, $\beta$=$t,b$=$1,2$ denote the top and bottom layers in each bilayer, respectively; $I,J$=$R,L$=$+1,-1$ denote the right and left bilayer systems (notice that if the positive $x$ direction is defined as indicated in the figure, their edge modes are right and left movers, respectively). An additional index $\mu=1,2$ is used to enumerate the two superconducting contacts within each layer. For example, the first superconducting contact on the top layer is denoted by $S_{\mu \alpha}$ with $\mu=1$ and $\alpha=t$ (or equivalently, $\alpha=1$). Similarly, the bosonic field describing the left moving mode on the bottom layer is denoted by $\phi_{I\alpha}$ with $I=L$ and $\alpha=b$, or alternatively $\phi_{-1,2}$.}
 \label{fig1}
\end{figure}

The difficulty of finding inherently non-Abelian states has recently spurred interest in artificial realizations of non-Abelian anyons using well established components, such as quantum wires, superconductors, topological insulators, and Abelian quantum Hall states as platforms. In particular, a tremendous amount of interest was invested in the construction of Majorana zero-modes from the interplay between spin-orbit coupling, superconductivity, and a magnetic field in one-dimensional systems or on the edges of two-dimensional systems \cite{Fu2008,Lutchyn2010,Oreg2010,Mourik2012,Das2012,Deng2012,Churchill2013,Alicea2012,Beenakker2013}. Recently, realizations of zero-energy bound states with a richer non-Abelian braiding statistics were proposed by coupling the edges of various Abelian FQH states via different mechanisms \cite{Lindner2012, Clarke2013, Cheng2012, Vaezi2014, Mong2014, Vaezi2014a}. Alternatively, closely related states were proposed in clean one-dimensional systems in the presence of strong interactions \cite{Oreg2014,Klinovaja2014b,Klinovaja2014c}.

Various different experiments capable of detecting signatures of these defects were proposed \cite{Barkeshli2014, Lindner2012, Clarke2013,Cheng2012}. In particular, Josephson effect experiments naturally possess such imprints in the form of unusual periodicities, reminiscent of the non-local nature of the non-Abelian anyons. For example, if the edges of two $\nu=1$ IQH states are gapped using a superconductor-backscattering-superconductor (SBS) heterostructure, the Josephson current is known to be $4\pi$-periodic as a function of the phase difference between the two superconductors due to the Majorana zero modes residing on the interfaces.  Similarly, replacing the IQH states by Laughlin FQH states, the Josephson current is now $4\pi \frac{e}{e^\star}$-periodic, where $e^\star$ is the minimal fractional charge in the system\cite{Clarke2013,Cheng2012}. This fact can be seen as a consequence of the parafermion zero-modes residing on the domain walls.

In this paper, we study the non-Abelian zero modes resulting from the interplay between multi-layer Abelian FQH states and superconductivity. Specifically, we focus on multi-layer systems with strong intra- and inter-layer interactions, in the absence of inter layer tunneling. Such a situation is realized, for example, in experiments with two twisted graphene layers forming a Moir\'e pattern, where the mismatch induced between the layers suppresses tunneling \cite{Luican2011}.

As we will demonstrate, by coupling the edges of two such $N$-component FQH states in an SBS heterostructure, one ends up with $N$ zero modes satisfying a generalized parafermion algebra presented in Eq. (\ref{pf}). For a general multi-layer system, we find that the ground state degeneracy associated with this algebra is given by $2^N |\text{Det}(K)|$, where $K$ is the $K$-matrix \cite{Wen1990a} of the multi-layer quantum Hall states. However, fixing the parities in the two layers reduces the number of accessible states to 1. We note that in more general situations, where an array of $M$ SBS junctions are considered, the ground state degeneracy is given by $\left(2^N |\text{Det}(K)|\right)^{M-1}$.

 In a specific set of symmetric multi-layer systems, we show that the generalized parafermion bound states can be decomposed into multiplications of $N$ conventional parafermion modes, each of which is non-local and does not generate zero-energy states individually.
We show that only one of these will be associated with the charge mode of the underlying FQH state, while the rest result from neutral modes.

We then turn to study the Josephson effect in our system. Similar to single-layer FQH states, we find that a standard Josephson effect experiment produces a $4\pi \frac{e}{e^\star}$-periodicity, where $e^\star$ is now the minimal charge in the multi-layer system. However, the possibility of realizing more elaborate configurations in the multi-layer system allows us to propose a novel Josephson effect experiment which is capable of probing additional quantum numbers, related to the neutral degrees of freedom in our system.

We do so by placing different superconducting contacts on different layers. This configuration is richer than the standard SBS heterostructure, as we now have control over the various relative phases between any pair of superconducting contacts.  By studying the energy as a function of the various relative phases, we show that all the quantum numbers associated with the underlying FQH state, including those related to the neutral modes, can be probed. In particular, we propose two relatively simple configurations: the first is the standard configuration which probes the charge sector (see Fig. \ref{super}(a)), while the second isolates the signal associated with the neutral modes  (see Fig. \ref{super}(b)).

Additionally, we find that inter-layer interactions induce a non-local Josephson effect. In particular, by changing the superconducting phase between the two superconducting contacts within a single layer, a Josephson current flows through the superconductors in the other layers.

The outline of this manuscript is as follows: In Sec. \ref{sec2}, we focus on bilayer FQH states ($N=2$), and study the properties of the non-Abelian defects generated in an SBS heterostructure. In Sec. \ref{sec3}, we turn to demonstrate how the Josephson effect can be used to probe non-trivial signatures of the resulting generalized parafermions. In Sec. \ref{sec4}, we extend the arguments presented in the previous sections to $N$-layer FQH states. Finally, in Sec. \ref{sec5} we summarize our results and conclude the paper. Some technical details have been relegated to the appendices.

\section{Generalized parafermions in bilayer FQH states}\label{sec2}
In this section we demonstrate the emergence of generalized parafermions in bilayer FQH systems. Later, in Sec. \ref{sec4}, we extend our results to more general multi-layer systems.

The system we study in this section is composed of two identical bilayer FQH systems, whose edges can backscatter by virtue of their spatial proximity (see Fig. \ref{fig1}).
Throughout this work, we refer to the edge modes propagating in $+x$ ($-x$) direction as right (left) movers, and use the notation $R$ ($L$), or interchangeably, $I=1$ ($I=-1$) to denote them.  Assuming the FQH states are Abelian with a single edge mode per layer, the most general low energy field theory describing the edge modes takes the chiral Luttinger liquid form  \cite{Wen1992}
\begin{equation}
S=\sum_{I=\pm1}\int\frac{dxdt}{4\pi}[I\cdot K_{\alpha\beta}\partial_t\phi_{I\alpha}\partial_x\phi_{I\beta}-V_{\alpha\beta}\partial_x\phi_{I\alpha}\partial_x\phi_{I\beta}].\label{eq3}
\end{equation}

Here, the fields $\phi_{I\alpha}$ are the different boson fields, enumerated by the chirality index $I$, and the layer indices $\alpha,\beta$ which take the values $t,b$ (where $t$ ($b$) represents the top (bottom) layer - see Fig. \ref{fig1}), or interchangeably, $\alpha,\beta=1,2$. In addition, $K$ is the $2\times2$ $K$-matrix. If the two layers are identical, it is explicitly given by
\begin{equation}
K_{\alpha\beta}=\left(
\begin{array}{cc}
      m &l \\
      l &m
    \end{array}
  \right).
\end{equation}
Since we focus on Fermionic systems we assume that $m$ $(l)$ is an odd (even) integer.
While the non-universal matrix $V$ depends on various microscopic details such as the form of interactions and confining potential, it must generally be positive definite.

The commutation relations of the chiral bosonic fields are given by
\begin{equation}
[\phi_{I\alpha}(x),\phi_{I\beta}(x')]=iI\pi K^{-1}_{\alpha\beta}\text{sgn}(x-x')-\pi \sigma^y_{\alpha\beta}\label{cm1}
\end{equation}
\begin{equation}
[\phi_{R\alpha}(x),\phi_{L\beta}(x')]=i\pi K^{-1}_{\alpha\beta}-\pi \sigma^y_{\alpha\beta},
\label{eq6}
\end{equation}
where $\sigma^y$ is the second Pauli-matrix.
Within the bosonic edge theory, the electronic operators are given by vertex operators
\begin{eqnarray}
\Psi_{R\alpha}&\equiv&e^{iK_{\alpha\beta}\phi_{R\beta}}\\
\Psi_{L\alpha}&\equiv&e^{-iK_{\alpha\beta}\phi_{L\beta}},
\end{eqnarray}
where repeated indices are summed over. \par

So far, the edge theory did not include coupling between the right and left bilayer systems. In what follows, we will introduce two such terms. To be specific, we will consider the effects of backscattering terms, and terms arising from proximity to an $s$-wave superconductor. We note that even if all the layers have parallel spins, an $s$-wave superconductor with strong spin orbit coupling can give rise to processes in which a Cooper-pair from the superconductor split between the layers.

For the purpose of generating non-Abelian defects, we separate the edges into the three regions shown in Fig. \ref{fig1}: the first and last regions, denoted by the letter `S',  are gapped by proximity coupling to an $s$-wave superconductor (indicated by the light and dark gray regions), while middle region, denoted by `B', is gapped by backscattering terms (indicated by the purple region). We use the index $\mu=1,2$ to label the two S-regions (see Fig. \ref{fig1}). For later use, it will prove useful to study a configuration in which distinct superconducting contacts are placed on different layers. To be specific, the S-terms are generated using 4 superconducting contacts - 2 for each layer - denoted by $S_{\mu \alpha}$.

This SBS configuration described above is useful for two reasons: (i) Each interface will be shown to give rise to non-trivial protected non-Abelian zero modes, which constitute a generalization of the well studied parafermion zero-modes. (ii) The B-region may serve as a weak link between the different superconductors, prompting us to use the Josephson effect for detecting the generalized parafermion zero modes. As we will see in Sec. \ref{sec3}, the possibility of placing distinct superconducting contacts on different layers provides us with a large degree of tunability, which in turn allows us to probe all the quantum numbers associated with the generalized parafermions and underlying FQH state.

Due to the symmetry between the two layers, the most relevant superconducting term we can write takes the form
\begin{eqnarray}
H_\text{S}=\int dx \Delta(x)\sum_{\alpha}\Psi_{R\alpha}\Psi_{L\alpha}+h.c. \nonumber \\ =2\int dx\Delta(x)\sum_{\alpha}\cos[2K_{\alpha\beta}\theta_{\beta}],\label{SC}
\end{eqnarray}
  where $\Delta$ is the amplitude of the superconducting term, and we have defined
  \begin{equation}
\varphi_{\alpha}\equiv\frac{\phi_{R\alpha}+\phi_{L\alpha}}{2},\;\theta_{\alpha}\equiv\frac{\phi_{R\alpha}-\phi_{L\alpha}}{2}.\label{theta}
\end{equation}
 Similarly, we write the backscattering terms in the form
\begin{eqnarray}
H_\text{B}=\int dx t(x)\sum_{\alpha}\Psi^{\dagger}_{R\alpha}\Psi_{L\alpha}+h.c.\nonumber \\ =2\int dx t(x)\sum_{\alpha}\cos[2K_{\alpha\beta}\varphi_{\beta}],\label{BS}
\end{eqnarray}
\noindent
where $t$ is the backscattering amplitude.

Notice that due to the absence of direct tunneling between the top and bottom layers, the coupling between them is manifested only through the off-diagonal element of the $K$-matrix and the interaction matrix $V$.

 \par
The only non-vanishing commutation relations of the fields in Eq. (\ref{theta}) take the form
\begin{equation}
[\varphi_{\alpha}(x),\theta_{\beta}(x')]=i\pi K^{-1}_{\alpha\beta}\Theta(x-x'),\label{eq16}
\end{equation}
\begin{equation}
[\varphi_{\alpha}(x),\varphi_{\beta}(x')]=-\pi \sigma^y_{\alpha\beta}.\label{Klein}
\end{equation}
We note that the commutation relation presented in Eq. (\ref{Klein}) results from Klein factors.

In what follows, we assume $\Delta$ and $t$ are negative and large enough such that $\theta_{\alpha}$ or $\varphi_{\alpha}$ are pinned to one of the minima of cosine potential in the corresponding regions:
\begin{eqnarray}
\text{S regions:  } K\vec{\theta}&=&\pi\vec{n}_S\label{eq20}\\
\text{B region: }K\vec{\varphi}&=&\pi\vec{n}_B,\label{eq21}
\end{eqnarray}
where
$\vec{\theta}\equiv(\theta_t,\theta_b)$, $\vec{\varphi}\equiv(\varphi_t,\varphi_b)$
and $\vec{n}_S$ and $\vec{n}_B$ are vectors of integer-valued operators. We can clearly describe our ground state manifold using the operators $n_{B,\alpha},n_{S,\alpha}$. However, we emphasize that due to Eq. (\ref{eq16}), these operators do not commute. In particular, the sectors with $\alpha\neq\beta$  are not decoupled.

We define following operators at the interfaces between the S and B regions:
\begin{eqnarray}
\gamma_{\alpha(\mu)}\equiv \lim_{\varepsilon\to+0}\exp\Bigl[i\Bigl(\theta_{\alpha}(x_{\mu}\mp\varepsilon)+\varphi_{\alpha}(x_{\mu}\pm\varepsilon)\Bigr)\Bigr].\notag\\
\;(\mu=1,2, \alpha=t,b) \label{eq:zero modes}
\end{eqnarray}
\noindent
Here, $\mu$ labels the interfaces between the corresponding S-regions and B (see Fig. \ref{fig1}), and $x_{\mu}$ is the position of the interface (we note that the upper (lower) signs correspond to $\mu=1 (2)$).

\begin{figure}[h]
 \begin{center}
  \includegraphics[width=80mm]{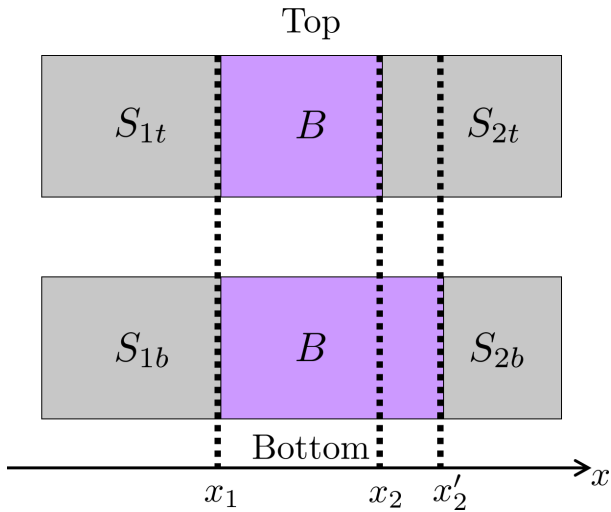}
 \end{center}
 \caption{An asymmetric configuration in which the B-region of the bottom layer overlaps with the $S_{2t}$ region in the top layer. We find that despite the asymmetry between the layers, the two zero-mode operators corresponding to different layers are located in the same position. Specifically, we find that depending on whether $\Delta$ or $t$ dominate, the generalized parafermions are located at $x_2$ or $x'_2$. This demonstrates the stability of the generalized parafermionic algebra.
 }
 \label{stable}
\end{figure}

These operators are zero modes as they commute with cosine terms appearing in Eqs. (\ref{SC}) and (\ref{BS}). This can most simply be seen by observing that they translate the arguments of the cosines by integer multiples of $2\pi$. Furthermore, using the commutation relations given in Eqs. (\ref{eq16}) and (\ref{Klein}), we obtain the algebra
\begin{eqnarray}
\gamma_{\alpha(1)}\gamma_{\beta(2)}&=&\gamma_{\beta(2)}\gamma_{\alpha(1)}e^{iK^{-1}_{\alpha\beta}\pi+\sigma^y_{\alpha\beta}\pi}.\notag\\
\gamma_{\alpha(\mu)}\gamma_{\beta(\mu)}&=&\gamma_{\beta(\mu)}\gamma_{\alpha(\mu)}e^{\sigma^y_{\alpha\beta}\pi}
\label{pf}
\end{eqnarray}
This relations constitute a generalized parafermionic algebra.

The algebra of the zero modes $\gamma_{\alpha(\mu)}$, together with the quantum numbers they carry imply that these are nothing but the projection of the local quasiparticle operators near the interfaces, onto the ground state manifold of the cosine terms.\par

It is illuminating to study the invariance of the algebra given in Eq. (\ref{pf}) with respect to asymmetry between the layers. For instance, we can argue that the generalized parafermionic algebra is robust by considering a configuration where the B region in the bottom layer is longer than its counterpart at the top layer, as portrayed in Fig. \ref{stable}. To be specific, the $\mu=2$ interface on the bottom (top) layer is located in $x_2$ ($x'_2$), with $x'_2>x_2$. In this geometry, there is an overlap between the B region at the bottom layer and the superconducting region $S_{2t}$ at the top layer. Naively, in this configuration it seems that the zero modes associated with the two layers are not located one above the other. As we discuss below, this is not the case, and the generalized parafermions are indeed robust to the asymmetry between the layers.

Due to the off-diagonal elements in the K-matrix, the arguments of the overlapping backscattering and superconducting terms do not commute. Therefore, the arguments of the two cosine terms cannot be pinned simultaneously. We independently study the two regimes in which the system is gapped:   First, if $t$ is large enough such that it flows to the strong coupling limit, $\Delta$ scales down to zero in the region $x_2<x<x'_2$. In the absence of a superconducting term, the gap caused by the backscattering term in the top layer now extends to $x'_2$. Clearly, the generalized parafermions will now be located at $x'_2$. If, on the other hand, $\Delta$ is large,  the backscattering term flows to zero in the region $x_2<x<x'_2$.  In this case, the generalized parafermions are found at $x_2$. The above arguments indicate that the two zero mode operators associated with the two layers are not spatially separated, making the generalized parafermionic algebra stable to asymmetry between the layers.

Based on the above, we find that our system supports localized generalized parafermion zero-modes, which satisfy richer exchange statistics compared to the parafermion modes found in single component (Laughlin-like) quantum Hall states \cite{Lindner2012, Clarke2013}. In what follows we calculate the ground state degeneracy resulting from the above generalized parafermionic algebra.

To do so, we construct eigenstates of the gauge invariant operator $\zeta _{\alpha} = \gamma_{\alpha(2)}^\dagger \gamma_{\alpha(1)}$.
\noindent
Since $[\zeta_t,\zeta_b]=0$, we can simultaneously diagonalize $\zeta_t$ and $\zeta_b$. We note that in the simple case in which the layers are decoupled and each is in an IQH state (i.e., the K-matrix is simply the identity matrix), the $\zeta_{\alpha}$ operators represent the parity of the two Majorana fermions in the corresponding layers. Similar to the Majorana case, the eigenstates of $\zeta_{\alpha}$ span the ground state manifold.

\begin{figure}[h]
 \begin{center}
  \includegraphics[width=75mm]{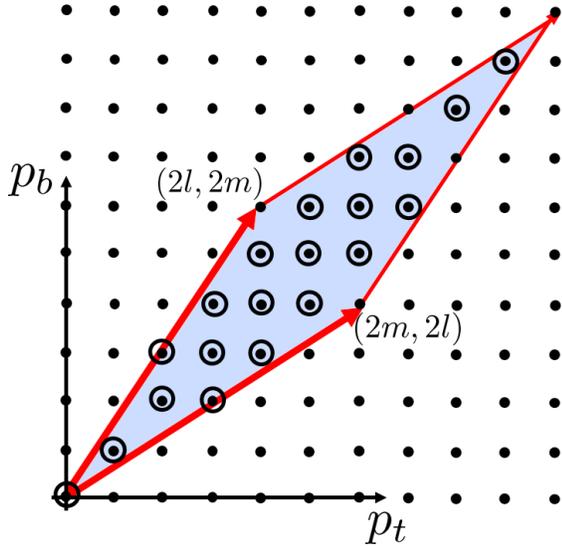}
 \end{center}
 \caption{The geometrical picture we use to show that the ground state degeneracy is given by $4\det(K)$ in the bilayer case. Here we set $m=3$ and $l=2$. Different points represent different $\vec{p}$-vectors, used to parameterize the eigenvalues of the $\zeta$-operators.  We end up with an enlarged lattice structure, whose unit cell (blue region) contains the $\vec{p}$-vectors which generate independent ground states. The corresponding independent ground states are denoted by circled lattice points. The area of the enlarged unit cell is equal to the ground state degeneracy.
}
 \label{det}
\end{figure}

To find the number of ground states which are simultaneous eigenstates of $\zeta_t$ and $\zeta_b$, we start by taking an arbitrary eigenstate $\left|\lambda_t,\lambda_b\right\rangle$, such that
\begin{equation}
\zeta_{\alpha}\left|\lambda_t,\lambda_b\right\rangle=\lambda_{\alpha}\left|\lambda_t,\lambda_b\right\rangle.
\end{equation}
We can now define a new set of ground states according to $(\gamma_{t(1)})^{p_t}(\gamma_{b(1)})^{p_b}\left|\lambda_t,\lambda_b\right\rangle$, where $p_b$ and $p_t$ are integers.

Using the generalized parafermionic algebra shown in Eq. (\ref{pf}), one finds that
\begin{eqnarray}
\zeta_{\alpha}(\gamma_{t(1)})^{p_t}(\gamma_{b(1)})^{p_b}\left|\lambda_t,\lambda_b\right\rangle \notag \\
=\lambda_{\alpha} e^{i\pi K^{-1}_{\alpha\beta}p_{\beta}}(\gamma_{t(1)})^{p_t}(\gamma_{b(1)})^{p_b}\left|\lambda_t,\lambda_b\right\rangle.\label{states}
\end{eqnarray}

We can therefore label the above states by \begin{equation}\left|\lambda_t e^{i\pi K^{-1}_{1\beta}p_{\beta}},\lambda_b e^{i\pi K^{-1}_{2\beta}p_{\beta}}\right\rangle=(\gamma_{t(1)})^{p_t}(\gamma_{b(1)})^{p_b}\left|\lambda_t,\lambda_b\right\rangle. \label{eq:sates multiplied by zeta}\end{equation} Assuming there are no additional quantum numbers describing the ground states, the number of distinct pairs of eigenvalues $\left(\lambda_t e^{i\pi K^{-1}_{1\beta}p_{\beta}},\lambda_b e^{i\pi K^{-1}_{2\beta}p_{\beta}}\right)$ gives us the ground state degeneracy.

\noindent

The form of the pairs of eigenvalues ensures that if $\vec{p}=(p_t,p_b)$ and $\vec{p'}=(p'_t,p'_b)$ are distinct two dimensional integer valued vectors related by $\vec{p'}=\vec{p}+2K\vec{q}$, where $\vec{q}$ is an integer valued vector, they represent identical ground states.
This equivalence relation can be visualized geometrically. First we note that the set of all integer valued vectors $\vec{p}$ form a two-dimensional square lattice (see Fig. \ref{det}). However, in our geometrical picture two lattice sites are considered equivalent only if they represent identical ground states. Therefore, we have a lattice structure with an enlarged unit cell, and the ground state degeneracy is given by the number of sites within a unit cell.

Noting that the unit cell of the underlying square lattice is of area 1, the number of elements within the enlarged unit cell is given by its area, which in turn is given by the area of parallelogram generated by the two primitive lattice vectors.

According to the equivalence relation $\vec{p}\sim\vec{p}+2K\vec{q}$, the two primitive lattice vectors are given by $2K\vec{e_1}=(2m,2l)^T$ and $2K\vec{e_2}=(2l,2m)^T$, where $\vec{e_1}=(1,0)^T$ and $\vec{e_1}=(0,1)^T$ (See Fig. \ref{det}). The area of the corresponding parallelogram is given by $4|\det[K]|$, showing that we get $4|\det[K]|$-fold ground state degeneracy. For example, in the simple case $m=3,l=2$, shown in Fig. \ref{det}, we get a 20-fold ground state degeneracy. As we will see, a similar result applies beyond the bilayer case: for $N$ layers, the ground state degeneracy is given by $2^N|\det[K]|$, where $K$ is now an $N\times N$ matrix.

However, assuming that the state of the bulk is given, and recalling that tunneling between the layers is absent, the parities $\zeta_\alpha$ are fixed. Working in such a physical sector, the degeneracy is reduced from $2^N|\det[K]|$ to 1. This is correct, however, only for a single SBS-junction. If an array of $M$ junctions is considered, only the total parities of the two layers, given by a multiplication of $\zeta_\alpha$ over the different junctions, are constrained. The degeneracy is therefore given by $\left(2^N|\det[K]|\right)^{M-1}$.

In the next section it will be useful to decompose the generalized parafermions into a multiplication of conventional parafermion operators, for which the $K$-matrix in Eq. (\ref{pf}) is diagonal. This is done by writing
\begin{eqnarray}
\gamma_{t(\mu)} & = & \eta_{c(\mu)}\eta_{n(\mu)},\nonumber \\
\gamma_{b(\mu)} & = & \eta_{c(\mu)}\eta_{n(\mu)}^{\dagger},\label{eq:generalized parafermions in terms of regular parafermions}
\end{eqnarray}
with
\begin{equation}
\eta_{c/n(\mu)}= \lim_{\varepsilon\to+0}\exp\Bigl[\frac{i}{2}\Bigl(\theta_{c/n}(x_{\mu}\mp\varepsilon)+\varphi_{c/n}(x_{\mu}\pm\varepsilon)\Bigr)\Bigr]\label{cn},
\end{equation}
and $\theta_{c/n}\equiv\theta_t\pm\theta_b$, $\varphi_{c/n}\equiv\varphi_t\pm\varphi_b$.
Notice that in the new basis, labeled by $c/n$, we have two independent sectors of the Hilbert space, generated by the two commuting canonical conjugate pairs, $\{\varphi_c,\theta_c\}$, $\{\varphi_n,\theta_n\}$. These correspond to the charged and neutral excitations, respectively.

The charge and neutral fields satisfy the commutation relations
\begin{eqnarray}
[\varphi_{c}(x),\theta_{c}(y)] & = & \frac{2\pi i}{m+l}\Theta(x-y)\\{}
[\varphi_{n}(x),\theta_{n}(y)] & = & \frac{2\pi i}{m-l}\Theta(x-y),
\end{eqnarray}
indicating that the operator $\eta_{c/n(\mu)}$ represent conventional parafermion operators, akin to the operators found in single-layer systems. It is, however, important to emphasize that while it is convenient to write the $\gamma$-operators in terms of the simpler $\eta$-operators, the latter are non-local and do not generate zero-energy modes. To be exact, if the $\eta$ operators act individually on the cosine terms in  Eqs. (\ref{SC}) and (\ref{BS}), they generate high energy excitations.

As we saw above, when two identical bilayer FQH systems are coupled via an SBS junction, we get generalized parafermion zero-modes.
In what follows, we devise a Josephson effect experiment which is capable of probing signatures of these modes.
\section{Imprints of the generalized parafermion modes using the Josephson effect}\label{sec3}
 We are now in a position to discuss the Josephson effect through the SBS heterostructure studied in the previous section. Similar to the single-component quantum Hall case, the periodicity of the Josephson effect is expected to provide imprints of the non-Abelian zero modes found at the interfaces. As we will show, the bilayer system provides us with a larger degree of tunability as the superconducting phases of each layer can in principle be controlled independently.

The relative phases between any two superconducting contacts can be controlled by connecting them with a superconducting wire, through which a magnetic flux can be threaded.
We denote the phase of the superconducting contact $S_{\mu \alpha}$ as $\delta_{\mu \alpha}$. We wish to calculate the dependence of the energy on the relative phases between the two superconductors in each layer. We note that in principle one should take the relative phases between different layers into account. However, as we exclude tunneling between the different layers, these will not affect the energy of the system (we note that the inter-layer interaction terms conserve the charge of the various layers, and are therefore not affected by such phases).

To understand how these physical phases enter the low energy theory described in Eqs. (\ref{eq3}), (\ref{SC}), and (\ref{BS}), it is desirable to make a connection between the microscopic degrees of freedom, in terms of which the physical phases are defined, to the low-energy ones. Such a connection is made possible within the coupled wires approach \cite{Kane2002,Teo2014,Klinovaja2013c,Seroussi2014,Neupert2014,Sagi2014,Klinovaja2014a,Meng2014,Santos2015,Sagi2015a,Gorohovsky2015,Meng2015,Mross2015,Meng2015a,Sagi2015,Meng2016,Isobe2015,Sahoo2015,Iadecola2016,Fuji2016,Huang2016}, in which one decomposes the double-layer system into an array of coupled wires. This approach generally provides us with analytically tractable microscopic models for fractional phases. In our case, we use the resulting microscopic model to demonstrate that the physical superconducting phases between the various contacts enter the low energy Hamiltonian as (see Appendix B)

\begin{align}
H_{S} & =\Delta\sum_{\alpha=t,b}\left[\int_{x<x_{\mu=1}}dx e^{i\delta_{1\alpha}}\Psi_{R\alpha}\Psi_{L\alpha}\right.\nonumber \\
 & \left.+\int_{x>x_{\mu=2}}dx e^{i\delta_{2\alpha}}\Psi_{R\alpha}\Psi_{L\alpha}\right]+h.c.,\label{eq40}
\end{align}
where $x_\mu$ denotes the location of the interface labeled by $\mu$.
It is useful to define the phase difference between the two superconducting phases on each layer as $\delta_{\alpha}\equiv\delta_{2\alpha}-\delta_{1\alpha}$.
Without affecting the energy, we can set $\delta_{1\alpha}=0$, and thus $\delta_{\alpha}=\delta_{2\alpha}$.

\begin{figure*}
 \begin{center}
\subfloat[]{
  \includegraphics[width=50mm]{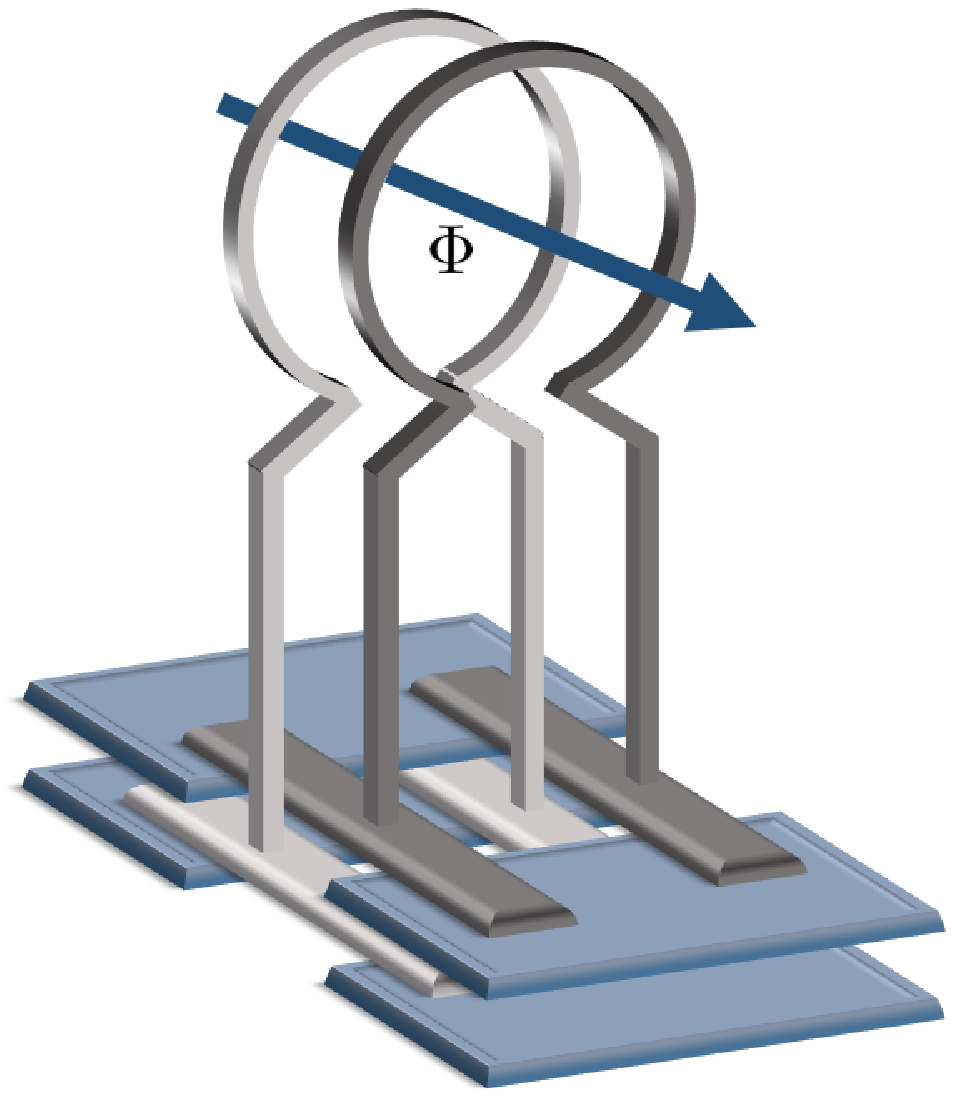}}
\subfloat[]{
  \includegraphics[width=50mm]{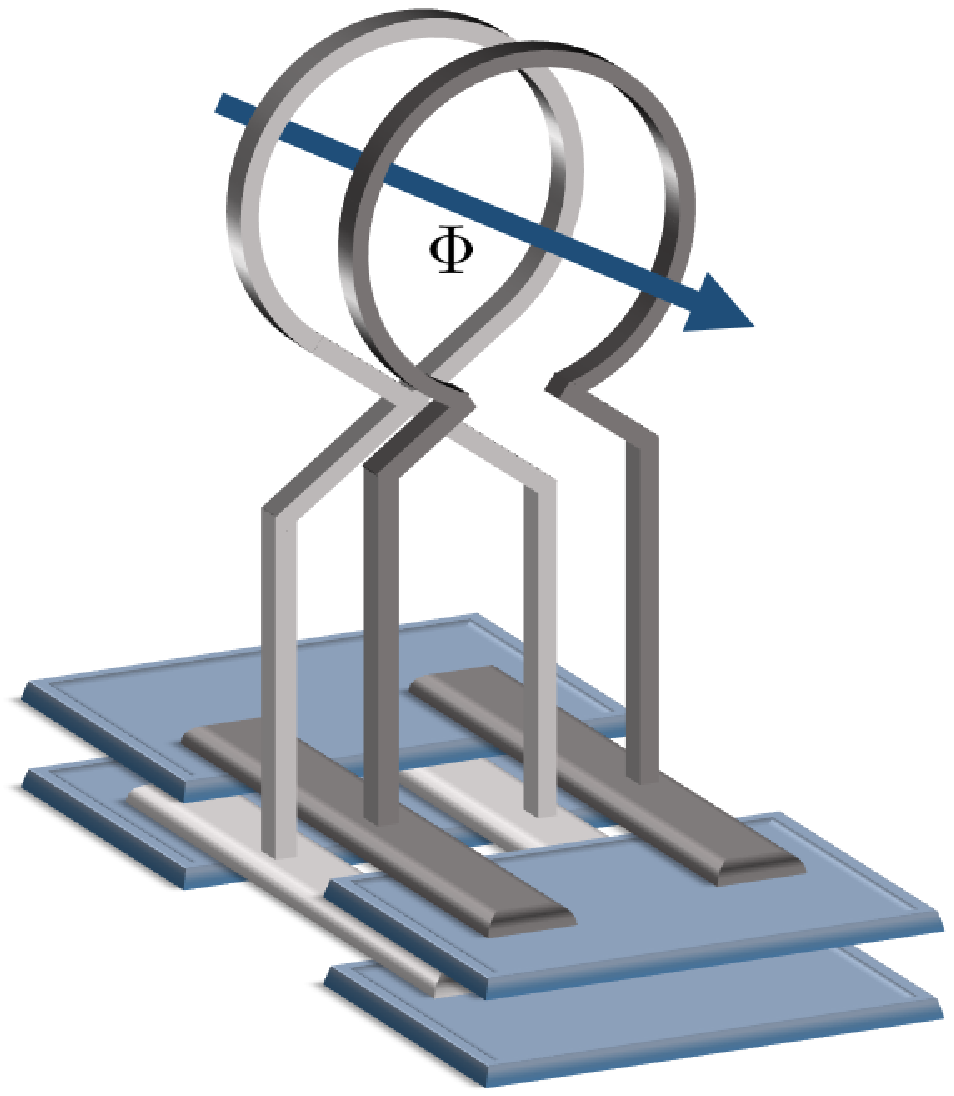}}
 \end{center}
 \caption{
In order to perform Josephson effect experiments, the two superconducting contacts on each layer are connected by a superconducting wire, through which a magnetic flux can be threaded. The magnetic flux controls the phase difference between the different superconductors.
In the configuration shown in (a), the phase differences in the top and bottom layers satisfy $\delta_t=\delta_b$. In this situation, we can only detect signatures of the charged degrees of freedom. In the configuration shown in (b), on the other hand, the phase differences in the top and bottom layers satisfy $\delta_t=-\delta_b$. This configuration isolates the signatures associated with the neutral degrees of freedom.}
 \label{super}
\end{figure*}

In terms of the bosonic degrees of freedom, Eq. (\ref{eq40}) takes the form
\begin{align}
H_{S} & =2\Delta\sum_{\alpha=t,b}\left[\int_{x<x_{\mu=1}}\cos[2K_{\alpha\beta}\theta_{\beta}]\right.\nonumber \\
 & \left.+\int_{x>x_{\mu=2}}\cos[2K_{\alpha\beta}\theta_{\beta}+\delta_{\alpha}]\right].\label{eqs2}
\end{align}
\noindent

The additional phases $\delta_{\alpha}$ may be eliminated by modifying the operators $\theta_{\alpha}$ in the region $x>x_{\mu=2}$ according to $\theta_{\alpha}\to\theta_{\alpha}+K^{-1}_{\alpha\beta}\frac{\delta_{\beta}}{2}$.
This, however, alters the generalized parafermion operators (Eq. \ref{eq:zero modes}) on the second interface.

The change of the generalized parafermionic operators affects the terms coupling different zero-mode operators. If such terms are considered, the ground state degeneracy is generally lifted. As we will see below, the energy of the resulting non-degenerate states now depends on the superconducting phases, leading to non-trivial Josephson currents $j_{\alpha}=\frac{\partial E_{GS}}{\partial \delta_{\alpha}}$ in the two layers.
To be explicit, the leading coupling terms are given by
\begin{equation}
H = \Gamma \left[\gamma^{\dagger}_{t(2)}\gamma_{t(1)}e^{-i\frac{K^{-1}_{1\beta}\delta_{\beta}}{2}}
+\gamma^{\dagger}_{b(2)}\gamma_{b(1)}e^{-i\frac{K^{-1}_{2\beta}\delta_{\beta}}{2}}\right]
+h.c.. \label{Eq: JE_Hamiltonian}
\end{equation}
As such terms are generated by tunneling of quasiparticles across the $B$-region,  the energy scale $\Gamma$ is proportional to the corresponding amplitude. We note that higher order terms, containing more than two $\gamma$-operators can also be considered. Such terms are, however, generated by tunneling of two or more quasiparticles, and are therefore expected to have smaller amplitudes. Moreover, they provide smaller periodicities.
In addition, notice that there are no coupling terms between zero-modes whose layer indices are different, as these are generated by quasiparticle tunneling between the layers, which is assumed to be absent.

The above low energy Hamiltonian has an explicit dependence on $\vec{\delta}=(\delta_t,\delta_b)$, from which we can find the Josephson currents. Notice that the Hamiltonian is written in terms of the commuting operators $\zeta_{\alpha}$, defined in the previous section. As these can be diagonalized simultaneously, we immediately write the energy spectrum as
\begin{equation}
E_{\vec{p}}(\vec{\delta}) = 2\Gamma \sum_{\alpha}\lambda_\alpha \cos\left[K^{-1}_{\alpha\beta}\left(\pi p_{\beta}- \frac{\delta_{\beta}}{2}\right)\right].
 \label{Eq: JE_energy}
\end{equation}

Remarkably, we find that even if we apply a phase difference between the superconducting contacts on one of the layers, we induce a Josephson current on the other layer due to the off-diagonal elements of K-matrix. This non-local Josephson effect is a direct consequence of the strong inter-layer interactions.

\begin{figure}[h]
 \begin{center}
  \includegraphics[width=88mm]{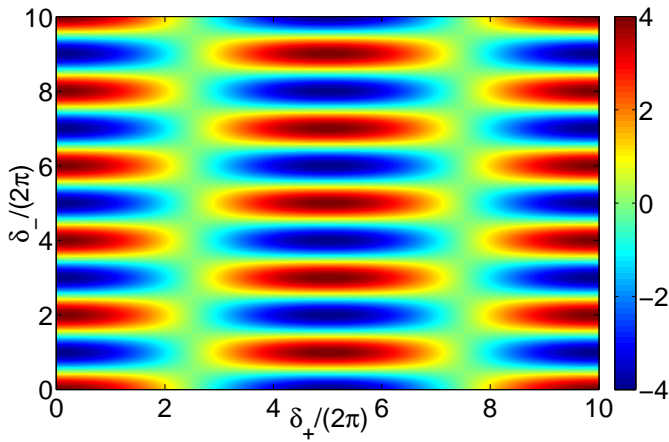}
 \end{center}
 \caption{\label{JE} The energy of the ground state characterized by $\vec{p}=0$ (in units of $\Gamma$) as a function of the two phases $\delta_{\pm}=\frac{\delta_t\pm\delta_b}{2}$ for the bilayer case with $m=3,l=2$. We find that the energy is $4(m+l)\pi=20\pi$-periodic ($4(m-l)\pi=4\pi$-periodic) as a function of the phase $\delta_+$ ($\delta_-$), which affects only the charge (neutral) sector of the theory.}
 \label{fig3}
\end{figure}

Taking, without loss of generality, $\lambda_\alpha =1$, we can write
\begin{equation}
E_{\vec{p}}(\vec{\delta})=4\Gamma\cos\left[\frac{2\pi p_{+}-\delta_{+}}{2\left(m+l\right)}\right]\cos\left[\frac{2\pi p_{-}-\delta_{-}}{2\left(m-l\right)}\right]\label{energy in terms of symmetric/antisymmetric phases}
\end{equation}
with $p_\pm =\frac{p_t\pm p_b}{2}$ and $\delta_\pm =\frac{\delta_t\pm \delta_b}{2}$.
In the simple case $m=3,l=2$, for example, we have 20 distinct vectors $\vec{p}$, as shown in Fig. \ref{det}. The dependence of $E_{\vec{p}=0}$ on the phases $\delta_{+}$ and $\delta_-$ is depicted in Fig. \ref{JE}.

It is evident that the above energy dependence shows non-trivial periodicities. To see this explicitly, it is helpful to examine two special situations:

(i) $\delta_t=\delta_b=\delta_{+}$: A realization of this situation is shown in Fig. \ref{super}(a). In this case, the energy becomes $4\Gamma_{-} \cos\left[\omega_+-\frac{\delta_{+}}{2(m+l)}\right]$, with $\Gamma_\pm=\Gamma\cos[\omega_\pm]$, and $\omega_{\pm}=\frac{\pi p_{\pm}}{m\pm l}$. By sweeping $\delta_{+}$, we obtain a $4(m+l)\pi$-periodic Josephson effect. Notice that $m+l$ can be written as $e/e^\star$, where $e^\star$ is the minimal fractional charge in the bilayer system, which is generally given by $e^\star/e =\vec{l}^T K^{-1} \vec{q}$ with $\vec{l}=\left(1,0\right)^T,\vec{q}=\left(1,1\right)^T$ for a symmetric double-layer system. This shows that the periodicity is given by $4\pi e/e^\star$.

(ii)$\delta_t=-\delta_b=\delta_{-}$: The configuration realizing this situation is illustrated in Fig. \ref{super}(b) (notice the opposite orientations of the two loops).
The energy now becomes $4\Gamma_+ \cos\left[\omega_--\frac{\delta_{-}}{2(m-l)}\right]$, leading to a $4(m-l)\pi$ periodicity as a function of $\delta_{-}$.

For the simple case $m=3$ and $l=2$, we obtain a $20\pi$-periodicity in $\delta_+$ and a $4\pi$-periodicity in $\delta_-$, as depicted in Fig. \ref{JE}.

 We now argue that the periodicities in situations (i) and (ii) can be interpreted as imprints of the charged and neutral degrees of freedom in our theory, respectively.
To see this, we first note that the superconducting terms in the regions $x>x_{\mu=2}$ can be written as
\begin{eqnarray}
4\Delta\int dx \cos\left[(m+l)\theta_c+\delta_+ \right]\cos\left[(m-l)\theta_n+\delta_-\right].\label{cosine in terms of charge neutral}
\end{eqnarray}

In case (i), we see that only the charge sector is affected by the introduction of the relative superconducting phases. It is therefore evident that the only component of $\gamma_{\alpha(\mu)}$ altered by $\delta_{+}$ is the operator $\eta_{c(\mu)}$, defined in Eqs. (\ref{eq:generalized parafermions in terms of regular parafermions}) and (\ref{cn}).
As the latter represents a regular parafermion operator associated with the charge sector, we find that the periodicity is given by $4\pi e/e^\star$. On the other hand, in case (ii) only the neutral sector is affected by the superconducting phase difference. We therefore find that $\delta_{-}$ alters the parafermion $\eta_{n(\mu)}$, associated with the neutral modes. Remarkably, configuration (ii), which is special to the bilayer case, enables us to detect signatures of the neutral modes. \par

We have shown in this section that the above two experiments isolate the signals associated with the charge and neutral sectors in our theory. In particular, by measuring the two types of periodicities studied above, one can measure all of the integers that characterize the generalized parafermion modes and bilayer FQH state (in our case, these are $l$ and $m$). As we will show in the next section, this remains correct for multi-layer FQH states with more than two layers.

\section{Generalization to $N$ layers}\label{sec4}
In this section, we generalize the analysis presented in the previous sections to general multi-layer FQH systems.
To do so, we prepare two copies of $N$-layer FQH states in the SBS configuration discussed above.

The low-energy description of the edges is given by Eq. (\ref{eq3}), where $K$ is now an $N\times N$ integer-valued $K$-matrix. Following the arguments presented in the previous sections, we get $N$ zero-mode operators $\gamma_{\alpha(\mu)}$ at the interfaces between the superconducting and backscattering regions. As expected, we obtain the same algebra as Eq. (\ref{pf}), where now the indices $\alpha,\beta$ take the values $\alpha,\beta=1,...,N$ and $\sigma^y_{\alpha,\beta}\rightarrow-i\text{sgn}(\beta-\alpha)$. Furthermore, following the geometric arguments presented for the double-layer case, we find that the ground state degeneracy generated by the generalized parafermionic algebra is generally given by $2^N|\det[K]|$. To be exact, working in a physical sector in which the parity in each layer is conserved, the degeneracy is given by $\left(2^N|\det[K]|\right)^{M-1}$, where $M$ is the number of SBS junctions.

As in the double-layer case, one can also study the imprints of the above generalized parafermions through the periodicity of the Josephson effect. As before, we consider a situation in which different superconducting contacts are placed on different layers, and the relative superconducting phases on the various layers are controlled independently.
Similar to the previous section, the terms coupling different zero-modes across the junction now take the form
\begin{equation}
H = \Gamma\sum_{\alpha=1}^{N} \left[\gamma^{\dagger}_{\alpha(2)}\gamma_{\alpha(1)}e^{-i\frac{\vec{e}_{\alpha}^T K^{-1}\vec{\delta}}{2}}\right]+h.c.,\label{cp}
\end{equation}
where $\vec{\delta}$ is now an $N$-dimensional vector containing the phase differences on the various layers, and the vectors ${{\vec{e}_\alpha}}$, whose elements are $(e_{\alpha})_{\beta}=\delta_{\alpha\beta}$, form an orthonormal basis in $N$-dimensional space.

In what follows, we focus on symmetric multi-layer systems, i.e., systems in which the $K$-matrix is unaffected by the exchange of any two layers. To be more specific, we take a $K$-matrix of the symmetric form
\begin{equation}
K_{\alpha\beta}=m\delta_{\alpha\beta}+l(1-\delta_{\alpha\beta}).\label{jain}
\end{equation}

If we first set $\delta_1=\delta_2=\cdots=\delta_{N}=\delta_+$, Eq. (\ref{cp}) becomes
\begin{equation}
H = \Gamma\sum_{\alpha=1}^{N} \left[\gamma^{\dagger}_{\alpha(2)}\gamma_{\alpha(1)}e^{-i\frac{\vec{e}_{\alpha}^T K^{-1}\vec{Q}}{2}\delta_+}\right]+h.c.,\label{cp2}
\end{equation}
where we have introduced the charge vector, $\vec{Q}=\left(1,1,...,1\right)^T$.

If the $N$-layers are symmetric, the expression $\vec{e_{\alpha}}K^{-1}\vec{Q}=1/(m+l(N-1))$ is independent of $\alpha$ and is generally given by $\frac{e^{\star}}{e}$, where $e^{\star}$ is the minimal charge in the multi-layer system. In this case, the periodicity of Josephson effect coincides with $4\pi\frac{e}{e^{\star}}$, as we saw in the single layer and bilayer cases.

Similar to the bilayer case, to detect imprints of the neutral modes, we choose the simple configuration $\delta_1=-\delta_2=\delta_-,\;\delta_3=\cdots=\delta_{N}=0$.
By varying $\delta_-$, we can find that the periodicity of Josephson effect is now given by $4(m-l)\pi$ (see Appendix \ref{ap3}). \par

To better understand the connection between the above periodicities and the underlying charge and neutral modes, one can decompose the generalized parafermion operators into a multiplication of conventional ones (see Appendix \ref{ap3}). Each conventional parafermion acts on the charge or one of the neutral sectors. We show in Appendix \ref{ap3} that the phase $\delta_+$ affects only the parafermion associated with the charge sector, while if $\sum_\alpha \delta_\alpha =0$, only the parafermions associated with the neutral modes are affected by the superconducting phases.

\section{Conclusions}\label{sec5}
In this work we study the emergence of generalized parafermions in multi-layer FQH states. We did so using the well-known SBS configuration, previously used to generate Majorana fermions or conventional parafermions in single layer systems.
We further examined the situation in which distinct superconducting contacts are placed on different layers. In this case, one can control the various relative phases, and independently measure the corresponding Josephson currents. This opens the door to novel Josephson effect experiments, through which various distinct properties of the generalized parafermions and the underlying multi-layer FQH states can be measured.

In particular, we demonstrated that the existence of off-diagonal $K$-matrix elements leads to a non-local Josephson effect in which the phase difference on one layer induces a Josephson current on the other layers. Furthermore, in symmetric multi-layer systems, we propose specific configurations in which the periodicity of the Josephson effect provides imprints of the neutral modes of the underlying FQH state, which are commonly inaccessible in Josephson effect experiments.

\section*{Acknowledgment}
This work was supported by Grants-in-Aid for Scientific Research on innovative areas ``Topological
Materials Science'' (Grants No. 15H0581, No. 15H0583) from the Ministry of Education, Culture,
Sports, Science, and Technology, Japan (MEXT), JSPS research fellowship (Grants No. 15J00565),
the Israel Science Foundation (ISF),
the European Research Council under the European Community's Seventh
Framework Program (FP7/2007-2013)/ERC Grant agreement No. 340210, DFG CRC TR 183, and the Adams Fellowship Program of the Israel Academy
of Sciences and Humanities.

\appendix

\section{A microscopic model using coupled-wires}\label{apb}
In this appendix, we present a coupled-wires construction of the multi-layer FQH states studied in main text. Using this microscopic model, we make a connection between the physical superconducting phases, defined in terms of the microscopic degrees of freedom, and those entering the low energy Hamiltonian studied in the main text. We will start by briefly reviewing the coupled-wire approach for single component FQH states. We will then generalize the approach to symmetric multi-layer FQH states.  Finally, we will use the resulting model to show that the phases applied between the superconducting contacts coincide with those appearing in the low energy Hamiltonian studied in the main text.

\subsection{IQHE states}

In the coupled-wires approach, we decompose the 2D quantum Hall system into an array of $M$ coupled
wires.

We work in a convention in which the wires are pointing in the $x$ direction and the magnetic field is pointing in the $z$ direction. If we choose a gauge in which the vector potential $\mathbf{A}$ also points
in the $\hat{x}$ direction, $k_x$ is a good quantum number. The energy spectrum as a function of $k_x$ is composed of shifted parabolas
(see  Refs. \cite{Kane2002,Teo2014}). To be specific, the energy spectrum of wire number $j$ is given by
\begin{equation}
E_j(k_x)=\frac{(k_x-2k_\phi j)^2}{2m},
\end{equation}
where we have defined $2k_{\phi}$
as the shift of two adjacent parabolas due to the magnetic field.
We define $k_{F}^{0}$ as the Fermi-momenta in the absence of an
external magnetic field .

Using bosonization, we write the low energy physics
in terms of $2M$ chiral boson fields $\Phi_{R/L}^{j}$,
such that
\begin{equation}
\psi_{R/L}^{j}\propto e^{i\left(\Phi_{R/L}^{j}+k_{R/L}^{j}x\right)},\label{eq:fermion operators}
\end{equation}
where $\psi_{R/L}^{j}$ is the fermion annihilation operator of the
right/left moving component of wire number $j$, $k_{R/L}^{j}$
is the corresponding Fermi-momentum.

Within this construction, the filling factor is given by
\begin{equation}
\nu=\frac{k_{F}^{0}}{k_{\phi}}.\label{eq:filling}
\end{equation}

Fixing the filling factor $\nu$, the Fermi momenta take the form
\begin{equation}
k_{R/L}^{j}=k_{\phi}\left(2j\pm\nu\right).\label{eq:momentum structure}
\end{equation}

In what follows we identify quantum Hall states by finding a set
of mutually commuting terms that conserve momentum, and can therefore acquire
an expectation value and completely gap out the bulk in the strong coupling limit. By an analysis
of the decoupled modes remaining on the edges, one is able to identify many QHE states \cite{Kane2002,Teo2014}.

The simplest case is $\nu=1$, which corresponds to the case where
adjacent parabolas cross at the chemical potential (i.e., $k_{R}^{j}=k_{L}^{j+1}$).
We see that simple tunneling operators between the wires conserve
momentum, and can therefore gap out the spectrum, leaving a single free chiral
mode on each edge. To be specific, the terms we consider are
\begin{equation}
\sum_{j=1}^{M-1}\cos\left(\Phi_{R}^{j}-\Phi_{L}^{j+1}\right).\label{eq:operators nu=00003D1}
\end{equation}
We see that the two modes $\phi_{L,1}$ and
$\phi_{R,M}$ remain free. These fields correspond to the chiral edge modes of
the $\nu=1$ case.

\subsection{Single component FQH states}

Using a similar approach, we can write models for the Laughlin states at filling
$\nu=1/m$, where $m$ is an odd integer.

As opposed to the integer
case, multi-electron processes are required in order to form a set
of commuting terms which conserve momentum. In order to identify this set of terms, it is convenient
to define the new set of chiral fields
\begin{equation}
\tilde{\Phi}_{R/L}^{j}=\frac{m+1}{2}\Phi_{R/L}^{j}+\frac{1-m}{2}\Phi_{L/R}^{j}.\label{eq:eta}
\end{equation}
The mapping is accompanied by a transformation of the momenta
\begin{equation}
q_{R/L}^{j}=\frac{m+1}{2}k_{R/L}^{j}+\frac{1-m}{2}k_{L/R}^{j}.\label{eq:q}
\end{equation}

It can now be checked that the transformed momenta
$q$ are satisfy the momentum structure presented in Eq (\ref{eq:momentum structure}) with $\nu=1$. We can therefore repeat the analysis of the $\nu=1$ case in terms of the transformed fields, and write operators
of the form
\begin{equation}
\sum_{j=1}^{M-1}\cos\left(\tilde{\Phi}_{R}^{j}-\tilde{\Phi}_{L}^{j+1}\right)\label{eq:operators nu=00003D1/3}
\end{equation}
 which conserve momentum and can gap out the bulk in the strong coupling limit.

 The above terms leave the two fields $\tilde{\Phi}^{1}_{L},\tilde{\Phi}^{M}_{R}$ decoupled. These fields, localized on the edge, satisfy the chiral
Luttinger liquid structure of the Laughlin-edge modes (notice that as the fields are localized on the edges, we omit the wire index):
\begin{equation}
\left[\tilde{\Phi}_{I}(x),\partial\tilde{\Phi}_{I}(x')\right]=2\pi imI\delta(x-x').\label{eq:commutation of etas}
\end{equation}
Here, we have introduced the index $I=R/L$, or interchangeably, $I=\pm1$.

To make contact with the conventional notations, we define the fields
\begin{equation}
\phi_{I}=I\frac{\tilde{\Phi}_I}{m},\label{eq:defining chi}
\end{equation}
which satisfy
\[
\left[\phi_{I}(x),\partial\phi_{I}(x')\right]=I\frac{2\pi i}{m}\delta(x-x').
\]
 In terms of these, the fermion operators take the form $\tilde{\psi}=e^{iI m\phi}$.

As we saw above, the analysis of the $\nu=1/3$ state was reduced to that of the $\nu=1$
state through the transformation (\ref{eq:eta}). In what follows we
turn to study multi-layer systems at fractional filling factors.
As we will see, the analysis of such systems can be reduced to the analysis of non-interacting multi-layer systems with  $\nu=N$.

\subsection{Symmetric multi-layer FQH states}

We now turn to study Abelian N-layer states,
described by an $N\times N$ $K$-matrix with
odd diagonal elements and even off diagonal elements. We focus on symmetric multi-layer systems with
 \begin{equation}K_{\alpha \beta} =m\delta_{\alpha \beta} +l (1-\delta_{\alpha \beta}).\label{Kmatrix}\end{equation} Within the coupled-wires approach, we therefore have $N$ layers, each containing  $M$ wires. For our purpose
we will focus on a multi-layer system, yet a similar analysis can be applied to
single layer hierarchical systems.

The charge vector $\mathbf{Q}$
in this case is a vector of dimension $N$ with all entries equal
1. The filling factor is generally given by
\begin{equation}
\nu=\mathbf{Q}^{T}K^{-1}\mathbf{Q}.\label{eq:nu}
\end{equation}
However, as this is the filling factor of the whole system, each layer
has a filling factor of the form $\nu_{\text{layer}}=\frac{1}{N}\mathbf{Q}^{T}K^{-1}\mathbf{Q}.$
We denote
the chiral boson fields as $\Phi_{R/L}^{j,l}$, where $j=1\ldots M$, and $l=1\ldots N$.

The associated Fermi momenta take the form
$k_{R/L}^{j,l}=k_{\phi}\left(2j\pm\nu_{\text{layer}}\right)$ (notice that the momenta are independent
of $l$). For convenience, we define $N$-dimensional vectors containing
the $N$ chiral fields for each value of $j$:
\begin{equation}
\mathbf{\overrightarrow{\Phi}}_{R/L}^{j}=\left(\begin{array}{c}
\Phi_{R/L}^{j,1}\\
\Phi_{R/L}^{j,2}\\
\vdots\\
\Phi_{R/L}^{j,N}
\end{array}\right).\label{eq:phi vector}
\end{equation}
To follow the same approach we used in the single layer case, we would like to map this state
to a multi-layer state whose filling is $\nu=N$.
This is done by a simple generalization of Eq. (\ref{eq:eta}):
\begin{equation}
\overrightarrow{\tilde{\Phi}}{}_{R/L}^{j}=\frac{K+1}{2}\overrightarrow{\Phi}{}_{R/L}^{j}+\frac{1-K}{2}\overrightarrow{\Phi}{}_{L/R}^{j}.\label{eq:eta vector}
\end{equation}
In terms of these fields the momentum structure indeed corresponds
to a multi-layer state with filling $N$. Therefore, terms of the form
\begin{equation}
\sum_{j}\cos\left(\tilde{\Phi}_{R}^{j,l}-\tilde{\Phi}_{L}^{j+1,l}\right)\label{eq:operators gapping eta vector}
\end{equation}
gap out the bulk in the strong coupling limit. The fields
$\overrightarrow{\tilde{\Phi}}{}_{L}^{1}$ and $\overrightarrow{\tilde{\Phi}}{}_{R}^{M}$
remain decoupled, and satisfy the commutation relations
\begin{align}
\left[\tilde{\Phi}_{I}^{\alpha}(x),\partial\tilde{\Phi}_{I}^{\beta}(x')\right] & =2i\pi I K_{\alpha\beta}\delta(x-x'),\label{eq:edge commutation of eta vectors}
\end{align}
with $I=R/L$ (or $\pm1$) (notice that we have again omitted the index $j$).
These are the low energy degrees of freedom defining the edge theory of our Abelian multi-layer state.

As the above fields all carry a charge of $1$, the $N$ electron
operators on the edge take the form
\begin{equation}
\Psi_{I\alpha}=e^{i\tilde{\Phi}_{I}^{\alpha}}.\label{eq:electron operator}
\end{equation}

To make contact with more conventional notations, we can define the
 fields $\phi$ as
\[
\phi_{I\alpha}=I K^{-1}_{\alpha\beta}\tilde{\Phi}_{I}^{\beta}.
\]
These fields satisfy the commutation relations
\[
\left[\phi_{I\alpha}(x),\partial\phi_{I\beta}(x')\right]=2i\pi I K_{\alpha\beta}^{-1}\delta(x-x'),
\]
similar to the low energy degrees of freedom used in the main text. In terms of these, the electron operator is
\[
\Psi_{I\alpha}=e^{iIK_{\alpha\beta}\phi_{I\beta}}.
\]

\subsection{The effects of the superconducting phases on the low-energy physics}

The terms we are writing in Sec. \ref{sec3}, in the presence of relative superconducting phases, take the general form
\[
\Delta\sum_{\alpha}\Psi_{R\alpha}\Psi_{L\alpha}e^{i\tilde{\delta}_{\alpha}}+h.c.
\]

We would like to find the connection between phases $\tilde{\delta}$
and the physical phases $\delta$ applied between the superconducting contacts connected
to the various layers (see Fig. \ref{super}). In terms of the microscopic
$\Phi$ edge degrees of freedom, which are indeed localized on the corresponding
layers, we can introduce the physical superconducting phases in the form
\[
\Phi{}_{L}^{1,\alpha}(x)\rightarrow\Phi{}_{L}^{1,\alpha}(x)+\delta_{\alpha}(x)/2
\]
\[
\Phi{}_{R}^{1,\alpha}(x)\rightarrow\Phi{}_{R}^{1,\alpha}(x)+\delta_{\alpha}(x)/2,
\]
in each superconducting region. Defining a vector $\vec{\delta}$
containing the various superconducting phases, and using the transformation shown in Eq.  (\ref{eq:eta vector}),
we get that the $\tilde{\Phi}$ fields transform as
\[
\vec{\tilde{\Phi}}\rightarrow\vec{\tilde{\Phi}}+\vec{\delta}/2.
\]

This shows that the physical phases $\delta_{\alpha}$ coincide
with the phases $\tilde{\delta}_{\alpha}$ appearing in the low energy Hamiltonian (Eq. (\ref{eq40})).

\section{Decomposition to charged and neutral degrees of freedom in the $N$-layer case}\label{ap3}
In this appendix, we decompose the zero mode operators $\gamma_{\alpha(\mu)}\;(\alpha=1,\cdots, N, \mu=1,2)$, found for symmetric $N$-layer systems in Sec. \ref{sec3}, into a multiplication of conventional parafermionic operators (i.e., operators satisfying the generalized parafermionic algebra shown in Eq. (\ref{pf}) with a diagonal $K$-matrix). Similar to the bilayer case, these will be associated with the charge and neutral degrees of freedom, respectively. To do this, we rewrite $\gamma_{\alpha(\mu)}$ as
\begin{eqnarray}
\gamma_{1(\mu)}&=&\eta_{c(\mu)}\eta_{n1(\mu)}\eta_{n2(\mu)}\cdots\eta_{nN-1(\mu)}\notag\\
\gamma_{2(\mu)}&=&\eta_{c(\mu)}\eta^{\dagger}_{n1(\mu)}\eta_{n2(\mu)}\cdots\eta_{nN-1(\mu)}\notag\\
\gamma_{3(\mu)}&=&\eta_{c(\mu)}\eta^{0}_{n1(\mu)}(\eta^{\dagger}_{n2(k)})^{2}\cdots\eta_{nN-1(\mu)}\notag\\
&\vdots&\notag\\
\gamma_{N(\mu)}&=&\eta_{c(\mu)}\eta^{0}_{n1(\mu)}\eta^{0}_{n2(\mu)}\cdots\left(\eta^{\dagger}_{nN-1(\mu)}\right)^{N-1},
\end{eqnarray}
where

\begin{align}
\eta_{c(\mu)} & =\lim_{\varepsilon\to+0}\exp\Bigl[\frac{i}{N}\Bigl(\theta_{c}(x_{\mu}\mp\varepsilon)+\varphi_{c}(x_{\mu}\pm\varepsilon)\Bigr)\Bigr],\nonumber \\
\eta_{nq(\mu)}= & \lim_{\varepsilon\to+0}\exp\Bigl[\frac{i}{(q+1)q}\Bigl(\theta_{nq}(x_{\mu}\mp\varepsilon)+\varphi_{nq}(x_{\mu}\pm\varepsilon)\Bigr)\Bigr],\nonumber \\
 & (q=1,\cdots,N-1)\label{eq:charge neutral parafermions}
\end{align}
with
\begin{equation}
\begin{pmatrix}\theta_{c}\\
\theta_{n1}\\
\theta_{n2}\\
\vdots\\
\theta_{n(N-1)}
\end{pmatrix}=U\begin{pmatrix}\theta_{1}\\
\theta_{2}\\
\theta_{3}\\
\vdots\\
\theta_{N}
\end{pmatrix};\begin{pmatrix}\varphi_{c}\\
\varphi_{n1}\\
\varphi_{n2}\\
\vdots\\
\varphi_{n(N-1)}
\end{pmatrix}=U\begin{pmatrix}\varphi_{1}\\
\varphi_{2}\\
\varphi_{3}\\
\vdots\\
\varphi_{N}
\end{pmatrix}\label{eq:transformation}
\end{equation}

and

\begin{equation}
U=\begin{pmatrix}1 & 1 & 1 & 1 & \cdots & 1\\
1 & -1 & 0 & 0 & \cdots & 0\\
1 & 1 & -2 & 0 & \cdots & 0\\
1 & 1 & 1 & -3 & \cdots & 0\\
\vdots & \vdots & \vdots & \vdots & \vdots & \vdots\\
1 & 1 & 1 & 1 & \cdots & -(N-1)
\end{pmatrix}.\label{eq:U}
\end{equation}

Written explicitly, the transformation takes the form:
\begin{align}
\theta_{c} & =\theta_{1}+\theta_{2}+\cdots+\theta_{N},\nonumber \\
\varphi_{c} & =\varphi_{1}+\varphi_{2}+\cdots+\varphi_{N},\nonumber \\
\theta_{nq} & =\theta_{1}+\cdots+\theta_{q}-q\theta_{q+1},\nonumber \\
\varphi_{nq} & =\varphi_{1}+\cdots+\varphi_{q}-q\varphi_{q+1}\nonumber \\
 & (q=1,\cdots,N-1).\label{eq:charge and neutral modes}
\end{align}
Recalling the commutation relations given in Eq. (\ref{eq16}), and noting that the transformation $U$ diagonalizes the $K$-matrix (Eq. (\ref{jain})), we find that the new fields satisfy the commutation relations
\begin{equation}
[\varphi_{c}(x),\theta_{c}(x')]=\frac{iN\pi}{m+(N-1)l}\Theta(x-x')
\end{equation}
\begin{equation}
[\varphi_{nq}(x),\theta_{nq}(x')]=\frac{iq(q+1)\pi}{m-l}\Theta(x-x')\;(q=1, \cdots, N-1)
\end{equation}

We therefore have $N$ commuting canonical conjugate fields, $\{\varphi_c, \theta_c\}$, $\{\varphi_{nq},\theta_{nq}\}$, generating independent sectors of the Hilbert space.  These sectors correspond to a single charge mode and $N-1$ neutral modes, respectively.
These results imply that $\eta_{c(k)}$ and $\eta_{nq(k)}$ are conventional parafermionic operators acting in the charge and various neutral sectors, respectively.
As in the bilayer case, these operators do not represent zero-modes as they generate high energy excitation of the cosine terms in Eq. (\ref{SC}) and (\ref{BS}).

Below, we will study the Josephson effect experiment discussed in the main text. We will show that the periodicity corresponding to the cases $\delta_1=\delta_2=\cdots=\delta_N=\delta_+$ and $\sum_{\alpha}\delta_{\alpha}=0$ can be interpreted as imprints of charge and neutral sectors respectively. \par

From Eq. (\ref{eqs2}), the $\theta$-fields satisfy
\begin{equation}
K\vec{\theta}+\frac{1}{2}\vec{\delta}=\pi\vec{n}_s,
\end{equation}
in the superconducting regions, where $\vec{\delta}$ is a vector containing the superconducting phases in the various layers.

Acting with $U$ on this equation, we get
\begin{align}
 & \left(\begin{array}{cccc}
m+(N-1)l\\
 & m-l\\
 &  & \ddots\\
 &  &  & m-l
\end{array}\right)\left(\begin{array}{c}
\theta_{c}\\
\theta_{n1}\\
\vdots\\
\theta_{nN-1}
\end{array}\right)\nonumber \\
 & =\pi\left(\begin{array}{c}
n_{c}\\
n_{n1}\\
\vdots\\
n_{nN-1}
\end{array}\right)-\frac{1}{2}\left(\begin{array}{c}
\delta_{c}\\
\delta_{n1}\\
\vdots\\
\delta_{nN-1}
\end{array}\right)\label{Eq. for theta}
\end{align}
with
\begin{eqnarray}
n_c=n_{s1}+n_{s2}+\cdots+n_{sN},\notag\\
n_{nq}=n_{s1}+\cdots+n_{sq}-qn_{sq+1}, \notag\\
(q=1,\cdots, N-1),
\end{eqnarray}
and similarly for $\delta_c$ and $\delta_{nq}$:
\begin{eqnarray}
\delta_c=\delta_{1}+\delta_{2}+\cdots+\delta_{N},\notag\\
\delta_{nq}=\delta_{1}+\cdots+\delta_{q}-q\delta_{q+1},\notag\\
(q=1,\cdots, N-1).
\end{eqnarray}
We therefore obtain
\begin{eqnarray}
\theta_c&=&\frac{1}{m+(N-1)l}(n_c\pi-\frac{1}{2}\delta_c)\\
\theta_{nq}&=&\frac{1}{m-l}(n_{nq}\pi-\frac{1}{2}\delta_{nq}).
\end{eqnarray}

Focusing first on the situation in which $\delta_1=\delta_2=\cdots=\delta_N=\delta_+$, we find that $\delta_{nq}=0$ for all $q$, and $\delta_c=N\delta_+$. This indicates that the only component of $\gamma_{\alpha(\mu)}$ affected by $\delta_S$ is $\eta_{c(\mu)}$, corresponding to the charge sector. The periodicity of the Josephson current in $\delta_+$ is given by $4\pi\left(m+(N-1)l\right)$, which indeed corresponds to $4\pi e^\star/e$.

 In the case where $\sum_{\alpha}\delta_{\alpha}=0$, $\delta_c$ vanishes and only the neutral sectors are influenced by superconducting phase. For example, if we set $\delta_1=-\delta_2=\delta_-, \delta_3=\cdots=\delta_N=0$, the only non-vanishing phase is $\delta_{n1}=2\delta_-$, affecting $\eta_{n1(\mu)}$. The periodicity of the Josephson current as a function of $\delta_-$ is given by $4\pi\left(m-l\right)$.

 Similarly, taking $\delta_1=\delta_2=\cdots=\delta_q=\delta_-$, $\delta_{q+1}=-q\delta_-$, and $\delta_{q+2}=\delta_{q+3}=\cdots=\delta_N=0$, the only non-vanishing phase is $\delta_{nq}=q(q+1)\delta_-$. In this general case, the periodicity is again given by $4\pi\left(m-l\right)$.


%

\end{document}